\documentclass[11pt]{article} 
\usepackage{hyperref} 
\pdfoutput=1 
 
\begin{document} 
 
\title{Dynamics of Water Entry} 
 
\author{Tadd T. Truscott$^1$, Jeffrey M. Aristoff$^2$, \& \\ Alexandra H. Techet$^1$  \\ 
\vspace{1pt} \\
{\small $^1$Department of Mechanical Engineering, Massachusetts Institute of Technology}\\ {\small Cambridge, MA 02139 USA}\\
{\small $^2$Department of Mathematics, Massachusetts Institute of Technology} \\ {\small Cambridge, MA 02139 USA}} 
 
\maketitle 
 
 
\begin{abstract} 
The hydrodynamics associated with water-entry of spheres can be highly
variable with respect to the material and kinematic properties of the sphere.
This series of five fluid dynamics videos illustrates several subtle but
interesting variations that can be seen. The first series of videos contrasts
the nature of impact ($Fr = U_o/\sqrt{gd} = 5.15$) between a hydrophilic
(wetting angle of $\alpha$ = 60$^\circ$) and hydrophobic sphere ($\alpha$ =
120$^\circ$), and illustrates how surface coating can affect whether or not an
air cavity is formed; the views from the side and from above are synchronized
in time. The second video series illustrates how spin and surface treatments
can alter the splash and cavity formation following water entry. The spinning
sphere ($S = \omega r / U_o = 1.7$; $Fr = 5.15$) causes a wedge of fluid to be
drawn into the cavity due to the no-slip condition and follows a curved
trajectory. The non-spinning sphere ($Fr = 5.15$) has two distinct surface
treatments on the left and right hemispheres: the left hemisphere is
hydrophobic ($\alpha$ = 120$^\circ$) and the right hemisphere is hydrophilic
($\alpha$ = 60$^\circ$). Interestingly, the cavity formation for the
half-and-half sphere has many similarities to that of the spinning sphere
especially when viewed from above. The third video series compares two
millimetric nylon spheres impacting at slightly different impact speeds ($U_o =
40 cm/s$ and $U_o = 45 cm/s$); the faster sphere fully penetrates the free
surface, forming a cavity, whereas the slower sphere does not. The fourth
series shows the instability of an elongated water-entry cavity formed by a
millimetric steel sphere with a hydrophobic coating impacting at $U_o = 600
cm/s$. The elongated cavity forms multiple pinch-off points along its decent.
\end{abstract} 
 
 
\section{Information} 
 
 
Two sample videos are 
\href{http://ecommons.library.cornell.edu/bitstream/1813/11479/2/Truscott_Aristoff_GFM2008.mpg}{(Video 1 mpeg-1)} and 
\href{http://ecommons.library.cornell.edu/bitstream/1813/11479/1/Truscott_Aristoff_GFM2008.m2v}{(Video 2 mpeg-2)}.
\\
\par
\noindent
\section{References}
Aristoff, J.M. \& Bush, J.W.M. ``Water Entry of Small Hydrophobic Spheres.'' {\it J Fluid
Mech.} In press 2008. \\
\\
\noindent
Truscott, T.T. \& Techet, A.H. ``Water-Entry of Spinning Spheres.'' {\it J Fluid Mech.} Under Review 2008.\\
\\
\noindent
Aristoff, J.M., Truscott, T.T., Techet, A.H. \& Bush, J.W.M. ``The Life of a Water-Entry Cavity at Low Bond Number.'' {\it Physics of Fluids}, September 2008.\\
\\
\noindent
Truscott, T.T. \& Techet, A.H. ``Cavity Formation in the Wake of a Spinning Sphere Impacting the Free Surface," {\it Physics of Fluids}, v. 18, no. 9, p. 18, cover, and inset, September 2006.\\

\end{document}